
\input harvmac

\def\d{\partial}

\def\Slash#1{{#1}{\llap/}}
\def\slash#1{\not\!\!#1}
\def\hf{{1 \over 2}}

\def\l{\left}
\def\r{\right}
\def\z0{z_0}
\def\w0{w_0}
\def\v#1{\vec{#1}}
\def\P#1{{\cal P}_{#1}}
\def\sitarel#1#2{\mathrel{\mathop{\kern0pt #1}\limits_{#2}}}
\def\ie{{\it i.e.}}

\def\np#1#2#3{{ Nucl. Phys.} {\bf B#1}, #2 (19#3)}

\def\pln#1#2#3{{Phys. Lett.} {\bf B#1}, #2 (19#3)}
\def\plo#1#2#3{{ Phys. Lett.} {\bf #1B}, #2 (19#3)}
\def\pr#1#2#3{{ Phys. Rev.} {\bf D#1}, #2 (19#3)}

\def\cmp#1#2#3{{Comm. Math. Phys.} {\bf #1}, #2 (19#3) }
\def\ann#1#2#3{{ Ann. Phys.} {\bf #1}, #2 (19#3)}
\def\ptp#1#2#3{{ Prog. Theor. Phys.} {\bf #1}, #2 (19#3)}

\def\jhep#1#2#3{{JHEP} {\bf #1}, #2 (19#3)}
\def\atmp#1#2#3{{Adv. Theor. Math. Phys.} {\bf #1}, #2 (19#3)}
\def\hpt#1{{\tt hep-th/#1}}

\lref\Mal{
J. Maldacena,
``The Large N Limit of Superconformal Field Theories and Supergravity'',
\atmp{2}{231-252}{98}, \hpt{9711200}.
}
\lref\GKP{
S. S. Gubser, I. R. Klebanov and A. M. Polyakov,
``Gauge Theory Correlators from Non-Critical String Theory'',
\pln{428}{105-114}{98}, \hpt{9802109}.
}
\lref\Witten{
E. Witten,
``Anti de Sitter Space and Holography'',
\atmp{2}{253-291}{98}, \hpt{9802150}.
}
\lref\ArfVol{
 I. Ya. Aref'eva and I. V. Volovich,
``On Large $N$ Conformal Theories, Field Theories in Anti-De Sitter Space
 and Singletons'',
\hpt{9803028};
``On the Breaking of Conformal Symmetry in the AdS/CFT Correspondence'',
\pln{433}{49-55}{98}, \hpt{9804182}.
}
\lref\MucVisS{
W. M\"{u}ck and K. S. Viswanathan,
``Conformal Field Theory Correlators from Classical Scalar
Field Theory on ${\rm AdS}_{d+1}$'',
\pr{58}{041901}{98}, \hpt{9804035}.
}
\lref\FMMRt{
D. Z. Freedman, S. D. Mathur, A. Matusis and L. Rastelli,
``Correlation functions in the ${\rm CFT}_d/{\rm AdS}_{d+1}$ correspondence'',
\np{546}{96-118}{99}, \hpt{9804058}.
}
\lref\CNSS{
G. Chalmers, H. Nastase, K. Schalm and R. Siebelink,
``R-Current Correlators in ${\cal N}=4$ Super Yang-Mills Theory 
from Anti-de Sitter Supergravity'',
\np{540}{247-270}{99}, \hpt{9805105}.
}
\lref\HenSfe{
M. Henningson and K. Sfetsos,
``Spinors and the AdS/CFT correspondence'',
\pln{431}{63-68}{98}, \hpt{9803251}.
}
\lref\MucVisV{
W. M\"uck and K. S. Viswanathan,
``Conformal Field Theory Correlators from Classical Field Theory on 
Anti-de Sitter Space II. Vector and Spinor Fields'',
\pr{58}{106006}{98}, \hpt{9805145}.
}
\lref\GKPF{
A. M. Ghezelbash, K. Kaviani, P. Parvizi, and A. H. Fatollahi,
``Interacting Spinors-Scalars and AdS/CFT Correspondence'',
\pln{435}{921}{98}, \hpt{9805162}.
}
\lref\AFG{
C. E. Arutyunov and S. A. Frolov,
``On the Origin of Supergravity Boundary Terms in the AdS/CFT 
Correspondence'',
\np{544}{576-589}{99}, \hpt{9806216}.
}
\lref\LiuTseyG{
H. Liu and A. A. Tseytlin,
``$D=4$ Super Yang-Mills, $D=5$ Gauged Supergravity,
and $D=4$ Conformal Supergravity'',
\np{533}{88-108}{98}, \hpt{9804083}.
}
\lref\MucVisG{
W. M\"{u}ck and K. S. Viswanathan,
``The Graviton in the AdS-CFT correspondence: Solution via the Dirichlet
boundary value problem'',
\hpt{9810151}.
}
\lref\AFGt{
G. E. Arutyunov and S. A. Frolov,
``Three-Point Function of the Stress Energy Tensor 
in the AdS/CFT Correspondence'',
\hpt{9901121}.
}
\lref\Corley{
S. Corley,
``The Massless Gravitino and the AdS/CFT Correspondence'',
\pr{59}{086003}{99}, \hpt{9808184}.
}
\lref\Volo{
A. Volovich,
``Rarita-Schwinger Field in the AdS/CFT Correspondence'',
\jhep{9809}{022}{98}, \hpt{9809009}.
}
\lref\KosRyt{
A. S. Koshelev and O. A. Rytchkov,
``Note on the Massive Rarita-Schwinger Field in the AdS/CFT Correspondence'',
\pln{450}{368-376}{99}, \hpt{9812238}.
}
\lref\AFP{
G. E. Arutyunov and S. A. Frolov,
``Antisymmetric Tensor Field on ${\rm AdS}_5$'',
\pln{441}{173-177}{98}, \hpt{9807046}.
}
\lref\lYiP{
W. S. l'Yi,
``Generating Functionals of Correlation Functions of P-form Currents 
in AdS/CFT Correspondence'',
\hpt{9809132}.
}
\lref\lYiMP{
W. S. l'Yi,
``Correlators of Currents Corresponding to the Massive P-form Currents 
in AdS/CFT Correspondence'',
\pln{448}{218-226}{99}, \hpt{9811097}.
}
\lref\Poli{
A. Polishchuk,
``Massive Symmetric Tensor Field on AdS'',
\hpt{9905048}.
}
\lref\DHFSk{
E. D'Hoker, D. Z. Freedman, W. Skiba,
``Field Theory Tests for Correlators in the AdS/CFT Correspondence'',
\pr{59}{045008}{99}, \hpt{9807098}.
}
\lref\Seiberg{
S. Lee, S. Minwalla, M. Rangamani and N. Seiberg,
``Three-Point Functions of Chiral Operators in $D=4, {\cal N}=4$
SYM at Large $N$'',
\atmp{2}{697-718}{98}, \hpt{9806074}.
}
\lref\LiuTsey{
H. Liu and A. A. Tseytlin,
``On Four-Point Functions in the CFT/AdS Correspondence'',
\pr{59}{086002}{99}, \hpt{9807097}.
}
\lref\FMMR{
D. Z. Freedman, S. D. Mathur, A. Matusis and L. Rastelli,
``Comments on 4-point Functions in the CFT/AdS Correspondence'',
\pr{59}{086002}{99}, \hpt{9808006}.
}
\lref\BroGut{
J. H. Brodie and M. Gutperle,
``String Correction to Four Point Functions in the AdS/CFT Correspondence'',
\pln{445}{296-306}{99}, \hpt{9809067}.
}
\lref\DHFV{
E. D'Hoker and D. Z. Freedman,
``Gauge Boson Exchange in ${\rm AdS}_{d+1}$'',
\np{544}{612-632}{99}, \hpt{9809179}.
}
\lref\ChaSch{
G. Chalmers and K. Schalm,
``The Large $N_c$ Limit of Four-Point Functions 
in ${\cal N}=4$ SuperYang-Mills Theory from Anti-de Sitter Supergravity'',
\hpt{9810051}; 
``Holographic Normal Ordering and Multi-Particle States 
in the AdS/CFT Correspondence'',
\hpt{9901144}.
}
\lref\Liu{
H. Liu,
``Scattering in Anti-de Sitter Space and Operator Product Expansion'',
\hpt{9811152}.
}
\lref\DHFS{
E. D'Hoker and D. Z. Freedman,
``General Scalar Exchange in ${\rm AdS}_{d+1}$'',
\hpt{9811257}.
}
\lref\HFMMR{
E. D'Hoker, D. Z. Freedman, S. D. Mathur, A. Matusis and  L. Rastelli,
``Graviton and Gauge Boson Propagators in ${\rm AdS}_{d+1}$'',
\hpt{9902042};
``Graviton Exchange and Complete 4-Point Functions 
in the AdS/CFT Correspondence'', 
\hpt{9903196}.
}
\lref\HFR{
E. D'Hoker, D. Z. Freedman and  L. Rastelli,
``AdS/CFT 4-Point Functions: 
How to Succeed at $z$ Integrals without Really Trying'',
\hpt{9905049}.
}
\lref\KleWit{
I. R. Klebanov and E. Witten,
``AdS/CFT Correspondence and Symmetry Breaking'',
\hpt{9905104}.
}
\lref\AFG{
C. E. Arutyunov and S. A. Frolov,
``On the Origin of Supergravity Boundary Terms in the AdS/CFT 
Correspondence'',
\np{544}{576-589}{99}, \hpt{9806216}.
}
\lref\Henx{
M. Henneaux,
``Boundary Terms in the AdS/CFT Correspondence for Spinor Fields'',
\hpt{9902137}.
}
\lref\table{
I. S. Gradshteyn and I. M. Ryzhik; A. Jeffrey ed., 
``Table of Integrals, Series, and Products (5th ed.)'',
(Academic Press, 1994).
}
\lref\Erde{
A. Erd\'elyi,
``Higher Transcendental Functions, Vol. 1'',
{\it Bateman Manuscript Project,}
(McGraw-Hill Book Company, Inc., 1953).
}
\lref\Erdea{
reference \Erde, 
page 103, (35).
}
\lref\Erdeb{
reference \Erde, 
page 102, (21) and (20).
}
\lref\BtBScP{
C. Fronsdal, 
``Elementary Particles in a Curved Space II'',
\pr{10}{589}{74}\semi
C. P. Burgess and C. A. Lutken,
``Propagators and Effective Potentials in Anti-de Sitter Space'',
\plo{153}{137}{85}\semi
T. Inami and H. Ooguri,
``One Loop Effective Potential in Anti-de Sitter Space'',
\ptp{73}{1051}{85}.
}
\lref\SUSYAdS{
C. J. Burges, D. Z. Freedman, S. Davis, and G. W. Gibbons,
``Supersymmetry in Anti-de Sitter Space'',
\ann{167}{285}{86}.
}
\lref\LSpP{
B. Allen and C. A. Lutken,
``Spinor Two-Point Functions in Maximally Symmetric Spaces'',
\cmp{106}{201-210}{86}.
}
\lref\kim{
H. J. Kim, L. J. Romans and P. van Nieuwenhuizen,
``Mass spectrum of chiral ten-dimensional $N=2$ supergravity on $S^5$'',
\pr{32}{389-399}{85}.
}


\Title{                                \vbox{\hbox{UT-848}
                                             \hbox{KEK-TH-624}
                                             \hbox{\tt hep-th/9905130}} }
{\vbox{\centerline{
                    Spinor Exchange in AdS$_{d+1}$ 
}}}

\centerline{ Teruhiko Kawano }
\vskip .1in 
\centerline{\sl Department of Physics, University of Tokyo}
\centerline{\sl Hongo, Tokyo 113-0033, Japan}
\centerline{\tt kawano@hep-th.phys.s.u-tokyo.ac.jp }
\vskip .1in
\centerline{ and }
\vskip .1in
\centerline{ Kazumi Okuyama }
\vskip .1in
\centerline{\sl High Energy Accelerator Research Organization (KEK)}
\centerline{\sl Tsukuba, Ibaraki 305-0801, Japan}
\centerline{\tt kazumi@post.kek.jp }

\vskip .4in


We explicitly calculate a Witten diagram with general spinor field exchange
on $(d+1)$-dimensional Euclidean Anti-de Sitter space, 
which is necessary to evaluate four-point correlation functions with 
spinor fields when we make use of the AdS/CFT correspondence, 
especially in supersymmetric cases. We also show that the amplitude can be 
reduced to a scalar exchange amplitude. 
We discuss the operator product expansion of the dual conformal field theory
by interpreting the short distance expansion of the amplitude 
according to the AdS/CFT correspondence.

\vskip 0.3cm
{\it PACS}: 11.25.Hf, 11.25.Sq, 11.15.-q

{\it keywords}: AdS/CFT correspondence, Spinor field, 
4-point function 
\Date{}


\newsec{Introduction}
\seclab\Intro

Recently it has been conjectured \Mal\ that string/$M$ theory on 
$(d+1)$-dimensional Anti-de Sitter space ${\rm AdS}_{d+1}$ 
times an Einstein manifold is dual to the large
$N$ limit of $d$-dimensional conformal field theory ${\rm CFT}_d$ 
on the boundary of ${\rm AdS}_{d+1}$. According to this conjecture, 
the dual of type IIB string theory on ${\rm AdS}_5\times
{\bf S}^5$ is 
four-dimensional 
${\cal N}=4$ $SU(N)$ supersymmetric Yang-Mills theory.  
In \refs{\GKP,\Witten}, a more precise dictionary of the AdS/CFT
correspondence was proposed to be that the partition function of 
the boundary CFT with source terms is equal to the partition function of 
the string theory on ${\rm AdS}_{d+1}$ with fixed boundary values of fields. 
Since the 't Hooft coupling $g^2_{\rm YM}N$ is proportional to the fourth power 
of the ratio of the radius of ${\rm AdS}_{d+1}$ to the string scale, 
the string partition function at the leading order in the strong coupling 
expansion can be estimated by classical type IIB supergravity.
In a schematic form, the correspondence can be written  \Witten\ as
\eqn\GKPW{
\l\langle \exp\l(\int_{\del( AdS)}{\cal O}_{\phi}\phi_0\r) \r\rangle_{{\rm CFT}}
=\exp\Big(-I_{{\rm SUGRA}}(\phi_{\rm cl})\Big)
}
where $\phi_{\rm cl}$ denotes the solution of the equations of motion of 
the fields satisfying the Dirichlet boundary condition in the supergravity and 
the boundary value of $\phi_{\rm cl}$ is identified with the source
$\phi_0$ up to a conformal factor.
This ``GKP/W relation'' \GKPW\ has been checked for two- and three-point 
functions 
\refs{\GKP\Witten\ArfVol\MucVisS\FMMRt\CNSS\HenSfe\MucVisV\GKPF\AFG
\LiuTseyG\MucVisG\AFGt\Corley\Volo\KosRyt\AFP\lYiP\lYiMP\Poli\DHFSk
{--}\Seiberg} 
and precise agreement between the two sides of \GKPW\ was shown. 

It is interesting and important  
to check the duality mapping \GKPW\ for four-point functions 
\refs{\MucVisS,\LiuTsey\FMMR\BroGut\DHFV\ChaSch\Liu\DHFS\HFMMR{--}\HFR}. 
On the CFT side, 
the forms of  two- and three-point functions are determined 
by conformal symmetry, but  four-point functions 
are only determined  up to a function of the cross ratios,
so they contain more information about the dynamics of 
the theory. Although correlation functions of 
the four-dimensional strongly coupled CFT have not been understood very well,
the GKP/W relation may give us some clues to the problem.
In \refs{\LiuTsey,\FMMR}, the four-point function of operators ${\cal O}_{\phi}$,
${\cal O}_C$ corresponding to dilaton field $\phi$ and axion field $C$ 
was considered.  
It was found in \FMMR, and in subsequent papers \refs{\DHFV,\ \Liu{--}\HFMMR}, 
that logarithmic terms of the cross ratios appear in various diagrams 
contributing to scalar four-point functions. In particular, such terms were shown 
\HFMMR\ to exist in the four-point function of ${\cal O}_{\phi}$, ${\cal O}_C$ 
after summing all the contributing diagrams.
In \Liu, it has been argued that the logarithmic terms are due to the  
mixture of a single-trace operator and double-trace operators 
in the operator product expansion which correspond to the exchanged scalar field
and the two external scalar fields respectively,
and that the double-trace operators 
in the CFT correspond to two-particle bound states in the supergravity on AdS.
Since the new states have not fully been understood, 
it is important to study whether two-particle bound states contribute to 
the other correlation functions.

In this paper, as a model, we will consider the Yukawa theory on 
${\rm AdS}_{d+1}$: 
a field theory on ${\rm AdS}_{d+1}$ of spinors and scalars with Yukawa 
interaction. In particular, we will calculate a diagram with spinor exchange 
in the model. This is a substantial step forward from the calculation of 
bosonic field exchange diagrams \refs{\Liu{--}\HFMMR}.
From the GKP/W relation, this diagram on AdS may contribute to 
four-point correlation functions of two spinor and two scalar operators 
in CFT. Therefore, our diagram is needed to obtain four-point functions 
in CFTs with spinor fields, for example, in supersymmetric models. 
In this paper, we will show that the spinor exchange amplitude 
can be reduced to the scalar exchange amplitude and its derivative.
This allows us to confirm the structure of the spinor indices of the 
four-point function and explicitly determine the function of cross ratios, 
which cannot be determined solely from conformal invariance.

This paper is organized as follows. 
In section 2, using the Yukawa theory as a model, we will illustrate how 
the spinor exchange diagram appears as the Witten diagram in the GKP/W relation. 
In section 3, we will show the calculation of the spinor exchange diagram by 
relating it to the scalar exchange diagram and present our result explicitly.
Section 4 is devoted to discussion about the operator product expansion of 
the spinor exchange amplitude. In Appendices A and B, the derivations of 
the bulk-to-bulk propagator and the Witten propagator of spinor fields 
are explained, respectively. We have also included some useful formulae 
regarding the propagators, which are indicated in the text.

\newsec{Yukawa model on the AdS Space}
\seclab\Yukawa

In this section, we will consider spinor exchange diagrams of
four-point correlation functions in a Yukawa model on 
$(d+1)$-dimensional Euclidean AdS space. In this paper, we will use  
the metric in the Poincar\'e representation;  
$ds^2={1 \over {z_0}^2} dz^{\mu} dz^{\mu}$.
The corresponding vielbein $e^a_{\mu}$ is 
$e^a_{\mu}={1 \over z_0}\delta^a_{\mu}$.
The Dirac operator $\slash{D}$ for spinor fields 
$\psi(z)$ is 
\eqn\Dirac{\eqalign{
\slash{D}\psi&=
e^{\mu}_{a}\Gamma^a\left(\d_{\mu}+\hf\omega_{\mu}^{bc}\Omega_{bc}\right)\psi
\cr
&=\l(z_0\Gamma^{a}\d_{a}-{d \over 2}\Gamma_0\r)\psi,
\cr}
}
where $\omega_{\mu}^{bc}$ is the spin connection given by
$\omega^{ab}_{\mu}
={1 \over z_0}(\delta^a_0\delta^b_{\mu}-\delta^b_0\delta^a_{\mu})$.
The gamma matrices $\Gamma^{a}$ satisfy 
$\{\Gamma^a,\Gamma^b\}=2\delta^{ab}$, and $\Omega_{bc}$ is defined by
$\Omega^{ab}={1\over4}[\Gamma^a,\Gamma^b]$.

In the Yukawa model, we have a spinor field $\psi$ and a scalar field 
$\phi$ with Yukawa interaction.
The action $S$ is given by
\eqn\action{
\int_{M}\llap d^{d+1}x\sqrt{g}\l[\bar\psi\l(\slash{D}-m\r)\psi
+\hf\l(\nabla_{\mu}\phi\nabla^{\mu}\phi+M^2\phi^2\r)
+\lambda\phi\bar\psi\psi\r]
+G\int_{\d M_{\epsilon}}\llap d^d\v{x}\sqrt{h}\bar\psi\psi,
}
where $\d M_{\epsilon}$ denotes a regularized boundary of AdS, 
\ie\ $\z0=\epsilon$. $h_{ij}$ is the induced metric on the
surface $\d M_{\epsilon}$. After the completion of our calculation, 
we will take the regularization parameter $\epsilon$ to zero. 
This prescription has been discussed for spinor fields in \MucVisV.
In \HenSfe, it was noted that the surface term must be added to the bulk 
Lagrangian, and the normalization of the 
term was determined in \AFG\ and \Henx\ to be $G=1$.

Now we will consider the solution of the equations of motion
\eqn\eom{\eqalign{
&\l(\slash{D}-m\r)\psi
=\l(z_0\Gamma^{a}\d_{a}-{d \over 2}\Gamma_0-m\r)\psi=-\lambda\phi\psi,
\cr
&\bar\psi\l(-\overleftarrow{\slash{D}}-m\r)
=\l(z_0\d_{a}\bar\psi\Gamma^{a}-{d\over2}\bar\psi\Gamma_0+m\bar\psi\r)
=-\lambda\phi\bar\psi,
\cr
&\l(\Delta-M^2\r)\phi=\lambda\bar\psi\psi,
\cr}
}
from the action \action, 
where $\Delta={1\over\sqrt{g}}\d_{\mu}\sqrt{g}g^{\mu\nu}\d_{\nu}$,
imposed by the Dirichlet boundary condition on $\z0=\epsilon$ 
\eqn\Dbc{\psi(z)=\psi_{\epsilon}(\v{z}),
\qquad\bar\psi(z)=\bar\psi_{\epsilon}(\v{z}),
\qquad\phi(z)=\phi_{\epsilon}(\v{z}).
}
Recalling the discussion in \MucVisV, 
we can only choose the boundary condition on either of the left- or
right-handed `chiral' components of these fields, as we will see below soon.
For the free case $\lambda=0$, 
the solutions $\psi=\psi_{\epsilon}^{(0)}$, 
$\bar\psi=\bar\psi_{\epsilon}^{(0)}$ can be immediately
obtained \MucVisV\ by using the Modified Bessel function $K_{\nu}(z)$ 
\table\ 
\eqn\freesol{\eqalign{
&\psi_{\epsilon}^{(0)}(z)
=\int{d^dk\over(2\pi)^d}e^{-i\v{k}\cdot\v{z}}
{\l({\z0 \over \epsilon}\r)}^{{d+1\over2}}
\l[{K_{m+\hf}(k\z0) \over K_{m+\hf}(k\epsilon)}
-i{\Gamma^ik^i \over k}{K_{m-\hf}(k\z0) \over K_{m+\hf}(k\epsilon)}\r]
\psi^{-}_{\epsilon}(\v{k}),
\cr
&\bar\psi_{\epsilon}^{(0)}(z)
=\int{d^dk\over(2\pi)^d}e^{-i\v{k}\cdot\v{z}}
{\l({\z0 \over \epsilon}\r)}^{{d+1\over2}}
{\bar\psi}^{+}_{\epsilon}(\v{k})
\l[{K_{m+\hf}(k\z0) \over K_{m+\hf}(k\epsilon)}
+i{\Gamma^ik^i \over k}{K_{m-\hf}(k\z0) \over
K_{m+\hf}(k\epsilon)}\r],
\cr}
}
where $\psi^{-}_{\epsilon}$ is the `chiral' component of $\psi_{\epsilon}$
which is defined such that $\Gamma^0\psi^{-}_{\epsilon}=-\psi^{-}_{\epsilon}$.
Similarly for $\bar\psi_{\epsilon}$,
$\bar\psi^{+}_{\epsilon}\Gamma^0=\bar\psi^{+}_{\epsilon}$.
In the limit $\epsilon\rightarrow0$, we have  
\eqn\Wprop{\eqalign{
\psi(z)&=
-\int d^d\v{x}\, U(z^{\mu}-x^{\mu})
K_{m+{d\over2}+\hf}(z,\v{x})\psi^{-}_{0}(\v{x}),
\cr
\bar\psi(z)&=\int d^d\v{x}\, {\bar\psi}^{+}_{0}(\v{x})
K_{m+{d\over2}+\hf}(z,\v{x})U(z^{\mu}-x^{\mu}),
\cr}
}
where $U(z-x)$ was introduced to be
\eqn\Killing{
U(z-x)={\Gamma^{\mu}(z^{\mu}-x^{\mu}) \over \sqrt{\z0}}.
}
$K_{m+{d\over2}+\hf}(z,\v{x})$ is the Witten propagator \refs{\Witten, \FMMRt}
of a scalar field with mass $m$
\eqn\ScWProp{
K_{\Delta}(z,\v{x})=
{\Gamma(\Delta)\over\pi^{d\over2}\Gamma(\Delta-{d\over2})}
\l({\z0\over\z0^2+|\v{z}-\v{w}|}\r)^{\Delta}
}
with $\Delta=m+{d\over2}+\hf$.
The boundary spinors 
$\psi^{-}_{0}(\v{x})$ and ${\bar\psi}^{+}_{0}(\v{x})$
in the limit $\epsilon\rightarrow0$ were defined 
such that $\psi^{-}_{0}=\epsilon^{m-{d\over2}}\psi^{-}_{\epsilon}$ and 
${\bar\psi}^{+} _{0}=\epsilon^{m-{d\over2}}{\bar\psi}^{+} _{\epsilon}$.
Therefore, the Witten propagators $\Sigma_{m+{d\over2}+\hf}$, 
$\bar\Sigma_{m+{d\over2}+\hf}$ of the spinor fields can be seen 
\refs{\HenSfe,\MucVisV} to be
\eqn\WPspin{\eqalign{
\Sigma_{\Delta}(z,\v{x})
&=U(z^{\mu}-x^{\mu})K_{\Delta}(z,\v{x})\P{-},
\cr
\bar\Sigma_{\Delta}(z,\v{x})
&=\P{+}K_{\Delta}(z,\v{x})U(z^{\mu}-x^{\mu}),
\cr}
}
where $\P{\pm}$ are the `chiral' projection operators given 
by $\P{\pm}=\hf(1\pm\Gamma^0)$. 

For the interacting case $\ie\ \lambda\not=0$, it is convenient to
introduce the bulk-to-bulk `regularized' propagator defined by  
\eqn\Green{
\l(\slash{D}-m\r)S_{\epsilon}(z,w)
=S_{\epsilon}(z,w)\l(-\overleftarrow{\slash{D}}-m\r)
={1\over\sqrt{g}}\delta^{d+1}(z-w),
}
with the boundary condition
\eqn\bcG{
\psi^{(0)}_{\epsilon}(z)
=\int d^d\v{w} \epsilon^{-d}
S_{\epsilon}(z,w)|_{\w0=\epsilon}\Gamma^{0}\psi_{\epsilon}(\v{w}).
}
The explicit form of the bulk-to-bulk propagator 
$S_{\epsilon}(z,w)$ 
is discussed in Appendix A, although we do not need it in this paper, 
except when verifying that 
\eqn\barbeG{
\bar\psi^{(0)}_{\epsilon}(z)=
-\int d^d\v{w} \epsilon^{-d}
\bar\psi_{\epsilon}(\v{w})\Gamma^{0}S_{\epsilon}(w,z)|_{\w0=\epsilon}.
}
In the limit $\epsilon\rightarrow0$, $S_{\epsilon}(z,w)$ turns out to be
\eqn\propagator{
S(z,w)
=-\l[\slash{D}+\Gamma^0+m\r]\z0^{-\hf}
\l[G_{{d\over2}+m-\hf}(z,w)\P{-}+G_{{d\over2}+m+\hf}(z,w)\P{+}\r]\w0^{\hf},
} 
where $G_{{d\over2}+m\mp\hf}(z,w)$ is the bulk-to-bulk scalar propagator 
given \refs{\BtBScP, \SUSYAdS} by
\eqn\scalarprop{
G_{\triangle}(z,w)=
{1\over4\pi^{{d+1\over 2}}}{1\over (u+1)^{\triangle}}
{\Gamma({\triangle\over2})\Gamma({\triangle+1\over 2})\over \Gamma(\nu+1)}
F\left({\triangle\over2},{\triangle+1\over 2},\nu+1;{1\over (u+1)^2}\right)
}
with $\triangle=\triangle_{\mp}={d\over2}+m\mp\hf$ 
and $\nu=\triangle-{d\over2}$.
The variable $u$ is the chordal distance; $u={(z-w)^2\over 2\z0\w0}$.
The scalar propagator satisfies that
\eqn\sgreen{
(-\Delta+m^2)G_{\Delta}(z,w)={1\over\sqrt{g}}\delta(z,w).
}
The function $F(\alpha,\beta,\gamma;z)$ is a hypergeometric function\Erde.
The derivation of the propagator \propagator\ is given in Appendix A.

Using the propagator \propagator, the solution of the equations of
motion \eom\ is given by a set of recursion relations \MucVisV
\eqn\recursion{\eqalign{
&\psi(z)=\psi^{(0)}_{\epsilon}(z)
-\lambda\int d^{d+1}w\sqrt{g_{w}}S_{\epsilon}(z,w)\phi(w)\psi(w),
\cr
&\bar\psi(z)=\bar\psi^{(0)}_{\epsilon}(z)
-\lambda\int d^{d+1}w\sqrt{g_{w}}\bar\psi(w)\phi(w)S_{\epsilon}(w,z),
\cr
&\phi(z)=\int d^d\v{x}K_{\epsilon}(z,\v{x})\phi_{\epsilon}(\v{x})
-\lambda\int d^{d+1}w\sqrt{g_{w}} G_{\epsilon}(z,w)\bar\psi(w)\psi(w)
\cr
&~~~~~~=\phi^{(0)}_{\epsilon}(z)
-\lambda\int d^{d+1}w\sqrt{g_{w}} G_{\epsilon}(z,w)\bar\psi(w)\psi(w),
\cr}
}
where $K_{\epsilon}(z,\v{x})$ is the `regularized' Witten propagator
\refs{\FMMRt, \MucVisS} of the scalar field. 
$G_{\epsilon}(z,w)$ is the `regularized' bulk-to-bulk propagator
\MucVisS\ of it. 
In this paper, we will not use the explicit form of them.
Solving the equation \recursion\ recursively and substituting 
the solution into the action \action, we find 
the bulk action of the spinors vanishing and obtain 
$S=S_B+S_F$, where $S_B$ gives the two-point function of the scalar
field in the boundary CFT and 
\eqn\solution{\eqalign{
&S_F=G\int_{\d M_{\epsilon}}d^d\v{x}\sqrt{h}\bar\psi(\v{x})\psi(\v{x})
\cr
&=G\int_{\d M_{\epsilon}}d^d\v{x}\sqrt{h}\bar\psi^{(0)}_{\epsilon}(\v{x})
\psi^{(0)}_{\epsilon}(\v{x})
+2\lambda G\int_{M} d^{d+1}z\sqrt{g(z)}\,\bar\psi^{(0)}_{\epsilon}(z)
\phi^{(0)}_{\epsilon}(z)\psi^{(0)}_{\epsilon}(z)
\cr
&-2\lambda^2 G\int_{M} d^{d+1}z\sqrt{g(z)}d^{d+1}w\sqrt{g(w)}
\, \bar\psi^{(0)}_{\epsilon}(z)\psi^{(0)}_{\epsilon}(z)
G_{\epsilon}(z,w)\bar\psi^{(0)}_{\epsilon}(w)\psi^{(0)}_{\epsilon}(w)
\cr
&-2\lambda^2 G\int_{M} d^{d+1}z\sqrt{g(z)}d^{d+1}w\sqrt{g(w)}
\, \bar\psi^{(0)}_{\epsilon}(z)\phi^{(0)}_{\epsilon}(z)
S_{\epsilon}(z,w)\phi^{(0)}_{\epsilon}(w)\psi^{(0)}_{\epsilon}(w)
\cr
&+O(\lambda^3),
\cr}
}
with $\sqrt{h}=\epsilon^{-d}$.
Note that the terms included in $O(\lambda^3)$ have more than four
external legs. Here we can see in \solution\ that the first term gives 
the two-point function of the spinors in the boundary CFT \refs{\MucVisV, \HenSfe}
and that the second term gives the three-point function 
with two spinors and a scalar \GKPF.
As in the case for scalar fields \refs{\FMMRt, \MucVisS}, we should 
take the limit $\epsilon\rightarrow0$ after the calculation 
when we evaluate the two-point function \MucVisV. But, since it seems that 
there is nothing wrong with the exchange of the order for 
the other multi-point functions, we will take the limit 
$\epsilon\rightarrow0$ at the beginning of the calculation.
The third and fourth term 
contribute to four-point functions in the boundary 
CFT. The third term $S_{\psi\bar\psi\psi\bar\psi}$ has four legs of
the spinor fields and can be easily related to a scalar exchange amplitude 
with four legs of scalars
\eqn\fourspinor{\eqalign{
&S_{\psi\bar\psi\psi\bar\psi}
\cr
&\sitarel{\rightarrow}{\epsilon\rightarrow0}
-2\lambda^2 G\int_{M}\llap d^{d+1}z\sqrt{g(z)}d^{d+1}w\sqrt{g(w)}
\, \bar\psi^{(0)}_{\epsilon}(z)\psi^{(0)}_{\epsilon}(z)
G_{\epsilon}(z,w)\bar\psi^{(0)}_{\epsilon}(w)\psi^{(0)}_{\epsilon}(w)
\cr
&=-2\lambda^2 G\int d^d\v{x}_1d^d\v{x}_2d^d\v{x}_3d^d\v{x}_4\  
{\bar\psi}^{+}_{0}(\v{x}_1)\slash{\v{x}}_{12}\psi^{-}_{0}(\v{x}_2)\ 
{\bar\psi}^{+}_{0}(\v{x}_3)\slash{\v{x}}_{34}\psi^{-}_{0}(\v{x}_4)
\cr
&~~~~~~~~~~~~~~~\times 
I_{\Delta}(\v{x}_1,m+{d+1\over2};\v{x}_2,m+{d+1\over2}
;\v{x}_3,m+{d+1\over2};\v{x}_4,m+{d+1\over2})
\cr}
}
with $\v{x}_{12}=\v{x}_1-\v{x}_2$ and $\v{x}_{34}=\v{x}_3-\v{x}_4$,
where 
the scalar exchange amplitude 
$I_{\Delta}(\v{x}_1,\Delta_1;\v{x}_2,\Delta_2;\v{x}_3,\Delta_3;
\v{x}_4,\Delta_4)$ is given by
\eqn\fourboson{\eqalign{
&\int d^{d+1}z\sqrt{g(z)}d^{d+1}w\sqrt{g(w)}\, 
\cr
&~~~~~~~~\times K_{\Delta_1}(z,\v{x}_1)K_{\Delta_2}(z,\v{x}_2)
G_{\Delta}(z,w)K_{\Delta_3}(z,\v{x}_3)
K_{\Delta_4}(z,\v{x}_4)
\cr}
}
which has been calculated by Liu \Liu\ and by D'Hoker and Freedman \DHFS\ 
using general values for the conformal dimensions $\Delta_a$, 
but here with $\Delta_a=m+{d+1\over2}$ for $a=1,\cdots,4$.
The fourth term with two spinor- and two scalar-legs
\eqn\spinexchange{\eqalign{
&S_{\psi\phi\bar\psi\phi}
\cr
&\sitarel{\rightarrow}{\epsilon\rightarrow0}-2\lambda^2 G
\int d^{d+1}z\sqrt{g(z)}d^{d+1}w\sqrt{g(w)}\, 
\phi_{\epsilon}^{(0)}(z)\bar\psi^{0}_{\epsilon}(z)
S_{\epsilon}(z,w)\psi^{0}_{\epsilon}(w)\phi_{\epsilon}^{(0)}(w)
\cr
&=2\lambda^2 G\int d^d\v{x}_1d^d\v{x}_2d^d\v{x}_3d^d\v{x}_4\, 
\int d^{d+1}z\sqrt{g(z)}d^{d+1}w\sqrt{g(w)}\,
\phi_0(\v{x}_2)K_{\Delta_2}(z,\v{x}_2) 
\cr
&~~~~~~\times
{\bar\psi}^{+}_{0}(\v{x}_1)
\bar\Sigma_{m+{d\over2}+\hf}(z,\v{x}_1)S(z,w)\Sigma_{m+{d\over2}+\hf}(w,\v{x}_3)
\psi^{-}_{0}(\v{x}_3)K_{\Delta_4}(w,\v{x}_4)\phi_0(\v{x}_4)
\cr
&=-2\lambda^2 G\int d^d\v{x}_1d^d\v{x}_2d^d\v{x}_3d^d\v{x}_4\,
\phi_0(\v{x}_2) 
{\bar\psi}^{+}_{0}(\v{x}_1)A(\v{x}_1,\v{x}_2,\v{x}_3,\v{x}_4)
\psi^{-}_{0}(\v{x}_3)\phi_0(\v{x}_4)
\cr}
}
remains to be calculated. The evaluation of this term is 
the main purpose of this paper, and we will show that this term can also be 
related to the scalar exchange amplitude $I_{\Delta}$ in the next 
section.

\newsec{Spinor Exchange Amplitude}
\seclab\Main

In this section, we will evaluate the spinor exchange diagram
\eqn\Amplitude{\eqalign{
&A(\v{x}_1,\v{x}_2,\v{x}_3,\v{x}_4)
\cr
&=-\int d^{d+1}z\sqrt{g(z)}d^{d+1}w\sqrt{g(w)}\,
K_{\Delta_2}(z,\v{x}_2)
\cr
&~~~~~~~~~~~~~\times \bar\Sigma_{\Delta_1}(z,\v{x}_1)S(z,w)
\Sigma_{\Delta_3}(w,\v{x}_3)
K_{\Delta_4}(w,\v{x}_4)
\cr
&=-\int d^{d+1}z\sqrt{g(z)}d^{d+1}w\sqrt{g(w)}\,
K_{\Delta_1}(z,\v{x}_1)K_{\Delta_2}(z,\v{x}_2) 
\cr
&~~~~~~~~~~~\times \P{+}U(z-{x}_1)S(z,w)U(w-{x}_3)\P{-}
K_{\Delta_3}(w,\v{x}_3)K_{\Delta_4}(w,\v{x}_4) 
\cr}
}
with the general value of the weights $\Delta_i$ $(i=1,\cdots,4)$.
The translational invariance on the boundary implies that 
$A(\v{x}_1,\v{x}_2,\v{x}_3,\v{x}_4)=A(\v{x}_{13},\v{x}_{23},0,\v{x}_{43})$. 
Under the inversion $z^{\mu}\rightarrow \hat z^{\mu}=z^{\mu}/z^2$ and 
$\v{x}\rightarrow\hat{\v{x}}=\v{x}/|\v{x}|^2$,
the integration measure is invariant and 
\eqn\inversion{\eqalign{
K_{\Delta}(\hat{z},\hat{\v{x}})&=|\v{x}|^{2\Delta}K_{\Delta}(z,\v{x}),
\cr
\bar\Sigma_{\Delta}(\hat{z},\hat{\v{x}})
&=-{\slash{\v{x}}\over|\v{x}|^{-2\Delta+2}}\P{-}
K_{\Delta}(z,\v{x})U(z-x){\slash{z}\over|z|},
\cr
\Sigma_{\Delta}(\hat{w},0)&=c_{\Delta}{\slash{w}\over|w|}
\w0^{\Delta-\hf}\P{-},
\cr}
}
with $|z|=\sqrt{z^2}$,
where $c_{\Delta}={\Gamma(\Delta)\over\pi^{d\over2}\Gamma(\Delta-{d\over2})}$.
Substituting $\v{x}_{i3}={\v{x}'_{i3}\over|\v{x}'_{i3}|^2}=\hat{\v{x}'}_{i3}$ 
($i=1,2,4$) into $A(\v{x}_1,\v{x}_2,\v{x}_3,\v{x}_4)$ and 
using the above equations \inversion, we can see that
\eqn\invA{
A(\v{x}_1,\v{x}_2,\v{x}_3,\v{x}_4)
=c_{\Delta_3}{\slash{\v{x}}_{13}\over|\v{x}_{13}|^{2\Delta_1}
|\v{x}_{23}|^{2\Delta_2}|\v{x}_{43}|^{2\Delta_4}}
B(\v{x}'_{13},\v{x}'_{23},\v{x}'_{43})
}
where
\eqn\AmpB{\eqalign{
B(\v{x}_{1},\v{x}_{2},\v{x}_{4})
=\P{-}\int d^{d+1}z &\sqrt{g(z)}d^{d+1}w\sqrt{g(w)}\,
K_{\Delta_2}(z,\v{x}_2)K_{\Delta_1}(z,\v{x}_1)
\cr
&\times U(z-x_1){\slash{z}\over|z|}S(\hat{z},\hat{w})
{\slash{w}\over|w|}
\w0^{\Delta_3-\hf}K_{\Delta_4}(w, \v{x}_4)\P{-}.
\cr}
}
As explained in Appendix A, $S(\hat{z},\hat{w})$ turns out to be
\eqn\invProp{
S(\hat{z},\hat{w})=
{\slash{z}\over|z|}\l(\slash{D}+\Gamma^0-m\r)
\l({1\over\z0}\r)^{\hf}
\l(G_{\Delta_{-}}(z,w)\P{+}+G_{\Delta_{+}}(z,w)\P{-}\r) \w0^{\hf}
{\slash{w}\over|w|},
}
where $\Delta_{\pm}={d\over2}+m\pm\hf$.
Putting \invProp\ into \AmpB, performing partial integration, and then 
using the formulas
\eqn\KUD{\eqalign{
&K_{\Delta}(z,\v{x})U(z-x)\overleftarrow{\slash{D}_z}
=-\l(\Delta-{d+1\over2}\r)\Gamma^0K_{\Delta}(z,\v{x})U(z-x),
\cr
&\z0\Gamma^{\mu}{\d\over\d z^{\mu}}K_{\Delta}(z,\v{x})
=\Delta\Gamma^0K_{\Delta}(z,\v{x})
-2\l(\Delta-{d\over2}\r)\l(\slash{z}-\slash{x}\r)K_{\Delta+1}(z,\v{x}),
\cr}
}
we obtain 
\eqn\AmpBtoJ{
B(\v{x}_{1},\v{x}_{2},\v{x}_{4})=
\l(2\Delta_2-d\r)J(\v{x}_{1},\v{x}_{2},\v{x}_{4})
+\l(m+\Delta_{12}-{d\over2}+\hf\r)I(\v{x}_{1},\v{x}_{2},\v{x}_{4})
}
with $\Delta_{12}=\Delta_1-\Delta_2$, 
where
\eqn\AmpIJ{\eqalign{
I(\v{x}_{1},\v{x}_{2},\v{x}_{4})
=&\int d^{d+1}z \sqrt{g(z)}d^{d+1}w\sqrt{g(w)}\,
K_{\Delta_2}(z,\v{x}_2)
\cr
&~~~~~~~\times K_{\Delta_1}(z,\v{x}_1)
G_{\Delta_{+}}(z,w)\w0^{\Delta_3}K_{\Delta_4}(w, \v{x}_4),
\cr
J(\v{x}_{1},\v{x}_{2},\v{x}_{4})
=&\int d^{d+1}z \sqrt{g(z)}d^{d+1}w\sqrt{g(w)}\,
\P{-}U(z-x_1)U(z-x_2)\P{-}
\cr
&~~~~~~~~\times
K_{\Delta_2+1}(z,\v{x}_2)K_{\Delta_1}(z,\v{x}_1)
G_{\Delta_{+}}(z,w)\w0^{\Delta_3}K_{\Delta_4}(w, \v{x}_4).
\cr}
}
From the following equations
\eqn\PUUP{\eqalign{
&\P{-}U(z-x_1)U(z-x_2)\P{-}=\l[{(z-x_2)^2\over\z0}
-\slash{x}_{12}{\Gamma^i(z^i-x^i)\over\z0}\r]\P{-},
\cr
&{\Gamma^i(z^i-x^i)\over\z0}K_{\Delta+1}(z,\v{x})
={1\over2(\Delta-{d\over2})}\slash{\d}_xK_{\Delta}(z,\v{x}),
\cr}
}
we find that
\eqn\JtoI{
J(\v{x}_{1},\v{x}_{2},\v{x}_{4})=
\l({1\over2\Delta_2-d}\r)
\l[2\Delta_2-\slash{\v{x}}_{12}\slash{\d}_2\r]
I(\v{x}_{1},\v{x}_{2},\v{x}_{4}).
}
Thus, we can obtain
\eqn\BtoI{
B(\v{x}_{1},\v{x}_{2},\v{x}_{4})=
\l[-\slash{\v{x}}_{12}\slash{\d}_2+\l(\Sigma_{12}+\Delta_{+}-d\r)\r]
I(\v{x}_{1},\v{x}_{2},\v{x}_{4})
}
with $\Sigma_{12}=\Delta_1+\Delta_2$.
As seen in \Liu\ and \DHFS, it is obvious that 
$I(\v{x}_{1},\v{x}_{2},\v{x}_{4})$ is essentially a scalar exchange amplitude. 
In this paper, we will follow the method which has 
been given by Liu \Liu\ to give our amplitude 
$B(\v{x}_{1},\v{x}_{2},\v{x}_{4})$ by using the Mellin-Barnes 
representation 
of the hypergeometric functions.
From the result of the paper \Liu, $I(\v{x}_{1},\v{x}_{2},\v{x}_{4})$ 
can be read to be
\eqn\AmpI{\eqalign{
&I(\v{x}_{1},\v{x}_{2},\v{x}_{4})
\cr
&={C\over2c_{\Delta_3}|\v{x}_{24}|^{\Sigma-2\Delta_3}}
\int^{{\Sigma_{12}\over2}+i\infty}_{{\Sigma_{12}\over2}-i\infty}{ds\over2{\pi}i}
\,\Gamma(-s+{\Sigma_{12}\over2})\Gamma(-s+{\Sigma_{34}\over2})
\Gamma(-s+{\Delta_{+}\over2})\,\tilde{I}(s)
\cr
&~~~~~~~~~~~~~~~~~~~~~~~~~~~~~~~~~~~~~~~~~\times\eta^{-s+{\Sigma_{12}\over2}}
F(s-{\Delta_{34}\over2},s+{\Delta_{12}\over2},2s;1-{\xi\over\eta}),
\cr}
}
where 
\eqn\coefC{\eqalign{
&C={1\over4\pi^{{3\over2}d}}
{\Gamma({\Sigma-d\over2})\Gamma({\Sigma_{12}+\Delta_{+}-d\over2})
\Gamma({\Sigma_{34}+\Delta_{+}-d\over2}) 
\over 
\Gamma(\Delta_{+}+1-{d\over2})\prod^{4}_{i=1}\Gamma(\Delta_i-{d\over2})},
\cr
&\tilde{I}(s)
={\Gamma({\Delta_{12}\over2}+s)\Gamma(-{\Delta_{12}\over2}+s)
\Gamma({\Delta_{34}\over2}+s)\Gamma(-{\Delta_{34}\over2}+s)
\over
\Gamma(2s)\Gamma({\Sigma+\Delta_{+}-d\over2}-s)}
\cr
&~~~~~~~~~~~~~~\times\ _3F_2\l({\widetilde{\Sigma}_{12}\over2},
{\widetilde{\Sigma}_{34}\over2},{\Delta_{+}\over2}-s;
\Delta_{+}+1-{d\over2},{\widetilde\Sigma\over2}-s;1\r),
\cr}
}
with $\widetilde{\Sigma}_{ij}=\Sigma_{ij}+\Delta_{+}-d$ and 
$\widetilde{\Sigma}=\Sigma+\Delta_{+}-d$. 
Additionally, $\Sigma_{ij}=\Delta_i+\Delta_j$, 
$\Delta_{ij}=\Delta_{i}-\Delta_{j}$, 
and $\Sigma=\sum^{4}_{i=1}\Delta_i$.
Here $\eta={|\v{x}_{14}|^2\over|\v{x}_{12}|^2}$ and 
$\xi={|\v{x}_{24}|^2\over|\v{x}_{12}|^2}$.
The function $\ _3F_2(a,b,c;e,f;z)$ is a generalized hypergeometric 
function \Erde. Therefore, putting \AmpI\ into \BtoI, we can immediately 
calculate $B(\v{x}_{1},\v{x}_{2},\v{x}_{4})$. Finally, from this 
$B(\v{x}_{1},\v{x}_{2},\v{x}_{4})$ and \invA, we obtain 
\eqn\result{
A(\v{x}_{1},\v{x}_{2},\v{x}_{3},\v{x}_{4})
=\Lambda(\v{x}_{1},\v{x}_{2},\v{x}_{3},\v{x}_{4})
\l[{\slash{\v{x}}_{13}\over|\v{x}_{13}|}A_1(\eta,\xi)
+{\slash{\v{x}}_{12}\over|\v{x}_{12}|}{\slash{\v{x}}_{24}\over|\v{x}_{24}|}
{\slash{\v{x}}_{43}\over|\v{x}_{43}|}A_2(\eta,\xi)\r],
}
where
\eqn\fcts{\eqalign{
&\Lambda(\v{x}_{1},\v{x}_{2},\v{x}_{3},\v{x}_{4})
={C\over|\v{x}_{12}|^{\Sigma_{12}}|\v{x}_{13}|^{\Delta_{12}-1}
|\v{x}_{23}|^{2\Sigma_{23}-\Sigma}|\v{x}_{24}|^{\Delta_{43}}
|\v{x}_{34}|^{\Sigma_{34}}},
\cr
&A_1(\eta,\xi)=
\int^{{\Sigma_{12}\over2}+i\infty}_{{\Sigma_{12}\over2}-i\infty}{ds\over2{\pi}i}
\,\Gamma(-s+{\Sigma_{12}\over2})\Gamma(-s+{\Sigma_{34}\over2})
\Gamma(-s+{\Delta_{+}\over2})\,\tilde{I}(s)
\cr
&~~~~~~~~~~~~~~~~~~\times\eta^{-s}(s+{\Delta_{+}-d\over2})
F(s-{\Delta_{34}\over2},s+{\Delta_{12}\over2},2s;1-{\xi\over\eta}),
\cr
&A_2(\eta,\xi)=
\int^{{\Sigma_{12}\over2}+i\infty}_{{\Sigma_{12}\over2}-i\infty}{ds\over2{\pi}i}
\,\Gamma(-s+{\Sigma_{12}\over2})\Gamma(-s+{\Sigma_{34}\over2})
\Gamma(-s+{\Delta_{+}\over2})\,\tilde{I}(s)
\cr
&~~~~~~~~~~~~~~~~~~\times\eta^{-s-\hf}
{(s-{\Delta_{12}\over2})(s-{\Delta_{34}\over2})\over2s}
F(s-{\Delta_{34}\over2},s+{\Delta_{12}\over2},2s+1;1-{\xi\over\eta}).
\cr}
}
Note that 
$\eta={|\v{x}_{13}|^2|\v{x}_{24}|^2\over|\v{x}_{12}|^2|\v{x}_{34}|^2}$
and that
$\xi={|\v{x}_{14}|^2|\v{x}_{23}|^2\over|\v{x}_{12}|^2|\v{x}_{34}|^2}$.
This is one of our main results in this paper.
As an easy check, we can see that 
the amplitude $A(\v{x}_{1},\v{x}_{2},\v{x}_{3},\v{x}_{4})$ consistently 
transforms under the inversion 
$\v{x}\rightarrow\hat{\v{x}}={\v{x}\over|\v{x}|^2}$,  
thanks to the structure of the gamma matrices 
$\slash{\v{x}}_{13}$ and 
$\slash{\v{x}}_{12}\slash{\v{x}}_{24}\slash{\v{x}}_{43}$.

\newsec{Discussion}
\seclab\Discs

In this paper, we have shown that the spinor exchange amplitude can be reduced 
to a scalar exchange amplitude and its derivative. By making use of this fact, 
we have calculated the spinor exchange amplitude explicitly.
From our final result \result\ and \fcts, we can see that 
three series of poles emerge from the three gamma functions, 
which is similar to the case of scalar fields \Liu.
as is similar to the case of scalar fields \Liu.
Therefore, we can easily see that the appearance of the logarithmic 
terms in the short distance expansion, for example 
when $\v{x}_{12}\rightarrow0$, has the same cause as in the scalar exchange 
diagram. Such terms appear because we have double poles in the integration 
over $s$ in \fcts\ when the position of the pole from one of the gamma functions 
coincides with that from another.

Apart from the issue of the logarithmic terms, we would like to discuss 
the operator product expansion in the CFT, according to the AdS/CFT 
correspondence. Let us suppose that we are now considering a Yukawa model 
with three spinor fields $\psi_1$, $\psi_3$, $\psi$ and two scalar fields 
$\phi_2$, $\phi_4$, which is the same situation as in section \Main.
The masses of the spinor fields $\psi_1$, $\psi_3$, and $\psi$ 
correspond to the weights $\Delta_1$, $\Delta_3$, and $\Delta_{+}$ 
respectively, and the scalar fields $\phi_2$, $\phi_4$ to the conformal 
dimensions $\Delta_2$, $\Delta_4$. The fields interact 
only through the following Yukawa interactions: 
$\lambda\phi_2\bar\psi_1\psi$ and $\lambda\phi_4\bar\psi\psi_3$.
In the rest of this section, we will show that the operator product expansions 
which can be determined from the two- and three-point functions are
consistent with at least the first two terms of 
the short distance expansion of the spinor exchange diagram 
which we calculated in the previous section, when we take account of 
only the contribution from the poles $s={\Delta_{+}\over2}$ 
in the exchange amplitude. 
For this purpose, let us recall the GKP/W relation \GKPW, in our case
\eqn\GKPWY{
\l\langle \exp\l(\int_{\del( AdS)}
\bar\psi^{+}_{0}\chi+\bar\chi\psi^{-}_{0}+{\cal O}\phi_0
\r) \r\rangle_{{\rm CFT}}
=\exp\Big(-I_{{\rm Yukawa}}(\phi_{\rm cl})\Big).
}
As suggested in section \Yukawa, the two-point correlation function 
of the spinors is given \refs{\HenSfe,\MucVisV} by
\eqn\twopoint{
\l\langle \chi_{\Delta}(\v{x}){\bar\chi}_{\Delta}(\v{y}) \r\rangle
=2c_{\Delta}{\Gamma^i(x^i-y^i)\over|\v{x}-\v{y}|^{2\Delta}}\P{-}.
}
Note that the weight $\Delta$ was $\Delta=m+{d+1\over2}$ 
in section \Yukawa. But, here, we do not assume any particular value for 
the conformal dimension, as we have mentioned above.
Similarly, we can find all the non-vanishing three-point functions \GKPF:
\eqn\threepoint{\eqalign{
&\l\langle \chi_{\Delta_1}(\v{x}_1){\cal O}_{\Delta_2}(\v{x}_2)
{\bar\chi}_{\Delta_{+}}(\v{x}_3) \r\rangle
=\lambda{2 C_{\Delta_1\Delta_2\Delta_{+}}\slash{\v{x}}_{13} 
\over |\v{x}_{12}|^{\Sigma_{12}-\Delta_{+}}
|\v{x}_{23}|^{\Delta_{+}-\Delta_{12}}|\v{x}_{31}|^{\Delta_{+}+\Delta_{12}}},
\cr
&\l\langle \chi_{\Delta_{+}}(\v{x}_1){\cal O}_{\Delta_4}(\v{x}_2)
{\bar\chi}_{\Delta_{3}}(\v{x}_3) \r\rangle
=\lambda{2 C_{\Delta_{+}\Delta_4\Delta_{3}}\slash{\v{x}}_{13} 
\over |\v{x}_{12}|^{\Delta_{+}-\Delta_{34}}
|\v{x}_{23}|^{\Sigma_{34}-\Delta_{+}}|\v{x}_{31}|^{\Delta_{+}+\Delta_{34}}},
\cr}
}
where the `structure constant' $C_{\Delta_1\Delta_2\Delta_3}$ is 
\eqn\strC{
C_{\Delta_1\Delta_2\Delta_3}
={\Gamma\l({\Delta_1+\Delta_2+\Delta_3-d\over2}\r)\over2\pi^{d}}
{\Gamma\l({\Sigma_{12}-\Delta_3\over2}\r)
\Gamma\l({\Sigma_{23}-\Delta_1\over2}\r)\Gamma\l({\Sigma_{31}-\Delta_2\over2}\r)
\over
\Gamma\l(\Delta_1-{d\over2}\r)\Gamma\l(\Delta_2-{d\over2}\r)
\Gamma\l(\Delta_3-{d\over2}\r)}.
}
Note that $C_{\Delta_1\Delta_2\Delta_3}$ is the structure constant of 
three-point functions of scalar fields with conformal dimensions 
$\Delta_1$, $\Delta_2$, and $\Delta_3$ \refs{\MucVisS,\FMMRt}. 
From this data \twopoint, \threepoint, 
we can determine the operator product expansions of the spinor field and 
the scalar field in the CFT, which turn out to be
\eqn\OPEc{\eqalign{
\chi_{\Delta_1}(\v{x}_1){\cal O}_{\Delta_2}(\v{x}_2)\sim
{{C_{\Delta_1\Delta_2}}^{\Delta_3}\over|\v{x}_{12}|^{\Sigma_{12}-\Delta_3}}
\chi_{\Delta_3}(\v{x}_2)
+{{D_{\Delta_1\Delta_2}}^{\Delta_3}\over|\v{x}_{12}|^{\Sigma_{12}-\Delta_3}}
\v{x}_{12}\cdot\v{\d}_2\chi_{\Delta_3}(\v{x}_2)
\cr
+{{S_{\Delta_1\Delta_2}}^{\Delta_3}\over|\v{x}_{12}|^{\Sigma_{12}-\Delta_3}}
\slash{\v{x}}_{12}\slash{\d}_2\chi_{\Delta_3}(\v{x}_2)+\cdots,
\cr}
}
\eqn\OPEa{\eqalign{
{\cal O}_{\Delta_4}(\v{x}_2)\bar\chi_{\Delta_3}(\v{x}_3)
\sim
\bar\chi_{\Delta_{+}}(\v{x}_2)
{{\bar C^{\Delta_{+}}}_{~~\Delta_4\Delta_3}\over
|\v{x}_{23}|^{\Sigma_{43}-\Delta_{+}}}
+\v{x}_{23}\cdot\v{\d}_2\bar\chi_{\Delta_{+}}(\v{x}_2)
{{\bar D^{\Delta_{+}}}_{~~\Delta_4\Delta_3}\over
|\v{x}_{23}|^{\Sigma_{43}-\Delta_{+}}}
\cr
+\bar\chi_{\Delta_{+}}(\v{x}_2)\overleftarrow{\slash{\d}}_2\slash{\v{x}}_{23}
{{\bar S^{\Delta_{+}}}_{~~\Delta_4\Delta_3}\over
|\v{x}_{23}|^{\Sigma_{43}-\Delta_{+}}}
+\cdots,
\cr}
}
where the coefficients are
\eqn\coefOPE{\eqalign{
&{C_{\Delta_1\Delta_2}}^{\Delta_3}={\bar C^{\Delta_3}}_{~~\Delta_2\Delta_1}
=\lambda{C_{\Delta_1\Delta_2\Delta_3} \over c_{\Delta_3}},
\cr
&{D_{\Delta_1\Delta_2}}^{\Delta_3}=-{\bar D^{\Delta_3}}_{~~\Delta_2\Delta_1}
=\lambda{\Sigma_{13}-\Delta_2\over2\Delta_3}
{C_{\Delta_1\Delta_2}}^{\Delta_3},
\cr
&{S_{\Delta_1\Delta_2}}^{\Delta_3}=-{\bar S^{\Delta_3}}_{~~\Delta_2\Delta_1}
=-\lambda{\Sigma_{23}-\Delta_1\over4\Delta_3
(\Delta_3-{d\over2})}{C_{\Delta_1\Delta_2}}^{\Delta_3}.
\cr}
}
It is convenient for later use to introduce coefficient $A_{\Delta}$ 
defined by $A_{\Delta}=\lambda^2{C_{12\Delta}C_{34\Delta}\over c_{\Delta}}$.
Now we are ready to consider the four-point function 
$\l\langle\chi_{\Delta_1}{\cal O}_{\Delta_2}\bar\chi_{\Delta_3}
{\cal O}_{\Delta_4}\r\rangle$.
By using the operator product expansions \OPEc, \OPEa, 
we can obtain the short distance expansion of the four-point function 
in the limit 
$\v{x}_{12}, \v{x}_{34} \rightarrow 0$, 
for terms lower than the order $O(\v{x}_{12}^2)$ or $O(\v{x}_{34}^2)$; 
\eqn\fourpoint{\eqalign{
&\l\langle\chi_{\Delta_1}(\v{x}_1){\cal O}_{\Delta_2}(\v{x}_2)
\bar\chi_{\Delta_3}(\v{x}_3){\cal O}_{\Delta_4}(\v{x}_4)\r\rangle
\cr
&\sim{A_{\Delta_{+}}\over|\v{x}_{12}|^{\Sigma_{12}-\Delta_{+}}
|\v{x}_{24}|^{2\Delta_{+}}
|\v{x}_{34}|^{\Sigma_{34}-\Delta_{+}}}\bigg[
\slash\v{x}_{13}-\l(\Delta_{+}+\Delta_{34}\r)(\v{x}_{43}\cdot\v{x}_{24})
{\slash\v{x}_{14}\over|\v{x}_{24}|^2}
\cr
&~-\l(\Delta_{+}+\Delta_{12}\r)(\v{x}_{12}\cdot\v{x}_{24})
{\slash\v{x}_{23}\over|\v{x}_{24}|^2}
-{(\Delta_{+}+\Delta_{12})(\Delta_{+}+\Delta_{34})\over2\Delta_{+}}
\bigg\{{(\v{x}_{12}\cdot\v{x}_{43})\over|\v{x}_{24}|^2}
\cr
&~-2(\Delta_{+}+1)
{(\v{x}_{12}\cdot\v{x}_{24})(\v{x}_{43}\cdot\v{x}_{24})\over|\v{x}_{24}|^4}
\bigg\}\slash\v{x}_{24}
+{(\Delta_{+}+\Delta_{21})(\Delta_{+}+\Delta_{43})\over4\Delta_{+}
(\Delta_{+}-{d\over2})}
{\slash\v{x}_{12}\slash\v{x}_{24}\slash\v{x}_{43}\over|\v{x}_{24}|^2}
\cr
&~~~+\cdots~~~~~~\bigg].
\cr}
}
This result is in complete agreement with 
the first two terms of the Taylor expansion of the spinor exchange amplitude 
$2\lambda^2 A(\v{x}_{1},\v{x}_{2},\v{x}_{3},\v{x}_{4})$
with respect to $\v{x}_{12}$, $\v{x}_{34}$, 
if we only take the contribution from the pole $s={\Delta_{+}\over2}$ 
into account and assume that 
${\Delta_{+}\over2}\not\in{\Sigma_{12}\over2}+{\bf Z}, 
{\Delta_{34}\over2}+{\bf Z}$: the condition that we do not have 
any logarithmic terms. On the other hand, under this condition, we have 
another contribution from the poles $s={\Sigma_{12}\over2}, {\Sigma_{34}\over2}$ 
and the subsequent poles. In the case that 
${\Sigma_{12}\over2}-{\Sigma_{34}\over2}\not\in{\bf Z}$, we do not have the 
logarithmic terms from those poles, either. Then we can see that 
the contribution from the pole $s={\Sigma_{12}\over2}$ does not agree with
the short distance expansion of the four-point function which can be obtained 
by combining two- and three-point functions in the same way as before, 
even though we do not assume any particular value for $c_{\Sigma_{12}}$ and 
$C_{\Delta_1\Delta_2\Sigma_{12}}$ in the two- and three-point functions, 
respectively. This discrepancy has been similarly observed in scalar exchange 
diagrams \Liu. 

If we do not impose any of the above condition on the weights, 
which is the case in the type IIB SUGRA on ${\rm AdS}_5\times{\bf S}^5$, 
we would have logarithmic terms in the spinor exchange diagram.
These terms have been discussed first in \FMMR\ and 
then in \refs{\BroGut\DHFV\ChaSch\Liu\DHFS{--}\HFMMR}.
Although these logarithmic terms had been expected to cancel each other 
in a realistic four-point function, the four-point function of scalar fields
was calculated in \HFMMR\ to show that such terms remain even in a realistic 
correlation function. Therefore it is likely that they also exist in realistic 
four-point functions with spinor fields.

\bigskip\bigskip
\centerline{{\bf Acknowledgements}}
We are grateful to Eric D'Hoker for discussion and for his encouragement.
T.K. was supported in part by Grant-in-Aid for Scientific Research
in a Priority Area: ``Supersymmetry and Unified Theory of Elementary 
Particles''(\#707) from the Ministry of Education, Science, Sports and 
Culture.
K.O. was supported in part through a grant from the JSPS Research Fellowship 
for Young Scientists.

\appendix{A}{The Bulk-to-Bulk Spinor Propagator}

In this appendix, we will give the derivation of the bulk-to-bulk 
propagator of spinor fields on the Euclidean ${\rm AdS}_{d+1}$ space 
(see \refs{\SUSYAdS, \LSpP} for the Lorentzian counterpart).
In addition, we will explain how the propagator transforms 
under the inversion, as we mentioned in the text.

Firstly, we seek the solution of the Dirac equation \eom\ with $\lambda=0$,
\ie 
\eqn\eomz{\eqalign{
&\l(\slash{D}-m\r)\psi
=\l(z_0\Gamma^{a}\d_{a}-{d \over 2}\Gamma_0-m\r)\psi=0,
\cr
&\bar\psi\l(-\overleftarrow{\slash{D}}-m\r)
=\l(z_0\d_{a}\bar\psi\Gamma^{a}-{d\over2}\bar\psi\Gamma_0+m\bar\psi\r)
=0.
\cr}
}
In momentum space, the solution is given by
\eqn\KIP{\eqalign{
&\bigg\{\matrix{
&\psi^{(K)}(\z0,\v{k})=\phi^{(K)}(\z0,\v{k})a^{-}(\v{k}),
\cr
&\psi^{(I)}(\z0,\v{k})=\phi^{(I)}(\z0,\v{k})b^{-}(\v{k}),
\cr}
\cr
&\bigg\{\matrix{
&\bar\psi^{(K)}(\z0,\v{k})
={\bar a}^{-}(\v{k})\bar\phi^{(K)}(\z0,\v{k})
={\bar a}^{-}(\v{k})\,i{\Slash{\v{k}} \over k}\,\phi^{(K)}(\z0,-\v{k}),
\cr
&\bar\psi^{(I)}(\z0,\v{k})
={\bar b}^{-}(\v{k})\bar\phi^{(I)}(\z0,\v{k})
=-{\bar b}^{-}(\v{k})\,i{\Slash{\v{k}} \over k}\,\phi^{(I)}(\z0,-\v{k}),
\cr}
\cr}
}
with $k=|\v{k}|$,
where $a^{-}(\v{k})$ and $b^{-}(\v{k})$ are functions of $\v{k}$ 
satisfying that $\Gamma^0(a^{-},b^{-})=-(a^{-},b^{-})$. 
Similarly for $\bar a^{-}(\v{k})$ and $\bar b^{-}(\v{k})$,  
$(\bar a^{-},\bar b^{-})\Gamma^0=-(\bar a^{-},\bar b^{-})$.
Additionally, 
\eqn\pfunc{\eqalign{
&\phi^{(K)}(\z0,\v{k})=\z0^{{d+1\over2}}
\l[K_{m+\hf}(k\z0)-i{\Slash{\v{k}} \over k}K_{m-\hf}(k\z0)\r],
\cr
&\phi^{(I)}(\z0,\v{k})=\z0^{{d+1\over2}}
\l[I_{m+\hf}(k\z0)+i{\Slash{\v{k}} \over k}I_{m-\hf}(k\z0)\r],
\cr}
}
where $K_{\nu}(z)$ and $I_{\nu}(z)$ are modified Bessel functions \table. 
For $|z|\rightarrow\infty$, the modified Bessel functions asymptotically 
behave \table\ as
\eqn\asymbesl{
K_{\nu}(z) \sim {\pi \over \sqrt{2z}} e^{-z}, \quad
I_{\nu}(z) \sim {e^z \over \sqrt{2\pi z}},
}
on the other hand, for $z\sim0$, 
\eqn\epsbesl{
K_{\nu}(z) \sim 2^{\nu-1}\Gamma(\nu)z^{-\nu},
\quad
I_{\nu}(z)\sim \l({z\over2}\r)^{\nu}{1\over\Gamma(\nu+1)}.
}

The propagator $S(z,w)$ should satisfy that
\eqn\green{
\l(\slash{D}-m\r)S(z,w)=S(z,w)\l(-\slash{D}-m\r)
={1\over\sqrt{g}}\delta^{d+1}(z-w)
}
with the regularity condition and the boundary condition
\eqn\regular{\eqalign{
&\lim_{\z0\rightarrow\infty}S(z,w)=\lim_{\w0\rightarrow\infty}S(z,w)=0,
\cr
&\lim_{\z0\rightarrow0}S(z,w)=\lim_{\w0\rightarrow0}S(z,w)=0.
\cr}
}
Since the modified Bessel functions satisfy the relation
\eqn\fmlbesl{
K_{m+\hf}(z)I_{m-\hf}(z)+I_{m+\hf}(z)K_{m-\hf}(z)={1 \over z},
}
we can verify that 
\eqn\propagator{\eqalign{
&S(z,w)
\cr
&=\int {d^d\v{k}\over(2\pi)^d} k 
e^{-i\v{k}\cdot(\v{z}-\v{w})}
\bigg[\theta(\z0-\w0)\phi^{(K)}(\z0,\v{k})\P{-}\bar\phi^{(I)}(\w0,-\v{k})
\cr
&~~~~~~~~~~~~~
-\theta(\w0-\z0)\phi^{(I)}(\z0,\v{k})\P{-}\bar\phi^{(K)}(\w0,-\v{k})\bigg]
\cr
\cr
&=-\z0^{-\hf}\l[\slash{D}+\hf\Gamma^0+m\r]
\l[G_{{d\over2}+m-\hf}(z,w)\P{-}+G_{{d\over2}+m+\hf}(z,w)\P{+}\r]\w0^{\hf}
\cr}
}
surely satisfies \green\ and \regular, if we use the expression of 
the bulk-to-bulk propagator \refs{\BtBScP, \SUSYAdS} of scalar fields
\eqn\sprop{\eqalign{
G_{\Delta}(z,w)=\int{d^d\v{k}\over(2\pi)^d}(\z0\w0)^{d\over2}
e^{-i\v{k}\cdot(\v{z}-\v{w})}
\bigg[&\theta(\z0-\w0)K_{\nu}(k\z0)I_{\nu}(k\w0)
\cr
&+\theta(\w0-\z0)K_{\nu}(k\w0)I_{\nu}(k\z0)\bigg]
\cr}
}
with $\Delta={d\over2}+\nu$, of which another expression can be seen in 
\scalarprop\ of the text. It will be useful later to differentiate 
the scalar propagator with respect to $u$:
\eqn\dprop{
{d\over du}G_{\Delta}(z,w)=G'_{\Delta}(z,w)=
-{\Delta C_{\Delta}\over(1+u)^{\Delta+1}}
F({\Delta\over2}+1,{\Delta+1\over2},\nu+1;{1\over(1+u)^2})
}
with $C_{\Delta}={\Gamma({\Delta\over2})
\Gamma({\Delta+1\over2})\over 4\pi^{{d+1\over 2}}\Gamma(\nu+1)}$.
By using some of the fifteen relations of Gauss on 
hypergeometric functions, we show that 
\eqn\GtoG{\eqalign{
(1+u)G'_{\Delta}(z,w)-(\Delta-d)G_{\Delta}(z,w)&=G'_{\Delta-1}(z,w),
\cr
(1+u)G'_{\Delta}(z,w)+\Delta G_{\Delta}(z,w)&=G'_{\Delta+1}(z,w).
\cr}
}
See \Erdea\ for the first formula and \Erdeb\ for the second.
From \dprop\ and \GtoG, an alternative expression of the spinor 
propagator $S(z,w)$ can be seen to be
\eqn\spinprop{
S(z,w)=-({1\over\z0\w0})^{\hf}
\l[\l(\slash{z}\P{-}-\P{+}\slash{w}\r)G'_{\Delta_{-}}(z,w)
+\l(\slash{z}\P{+}-\P{-}\slash{w}\r)G'_{\Delta_{+}}(z,w)\r],
}
where $\Delta_{\pm}={d\over2}+m\pm\hf$.

Now we proceed to consider the propagator $S(z,w)$ under 
the inversion transformation
$z^{\mu}\rightarrow\hat{z}^{\mu}={z^{\mu}\over z^2}$.
Since $G'_{\Delta_{\pm}}(z,w)$ is invariant under the inversion \inversion\ 
because of the invariance of the chordal distance $u$, it is easy to verify 
from the expression \spinprop\ that
\eqn\invspinprop{
S(\hat{z},\hat{w})=
{\slash{z}\over|z|}\l[\slash{D}+\Gamma^0-m\r]\z0^{-\hf}
\l[G_{\Delta_{+}}(z,w)\P{-}+G_{\Delta_{-}}(z,w)\P{+}\r]\w0^{\hf}
{\slash{w}\over|w|}.
}
Note that the middle factor in the right-hand side of \invspinprop\ is
$(-1)$ times the propagator \spinprop\ with mass $-m$ instead of $m$.

Finally, we will give the explicit form of the regularized bulk-to-bulk 
spinor propagator $S_{\epsilon}(z,w)$ defined by \Green\ and \bcG\ 
in the text, which satisfies the same differential equation as $S(z,w)$ 
but with the boundary condition \bcG. By making use of 
the solution \pfunc, we obtain
\eqn\regprop{\eqalign{
&S_{\epsilon}(z,w) 
\cr
&= S(z,w)
+\int {d^d\v{k}\over(2\pi)^d} k e^{-i\v{k}\cdot(\v{z}-\v{w})}
{I_{m+\hf}(k\epsilon)\over K_{m+\hf}(k\epsilon)}
\phi^{(K)}(\z0,\v{k})\P{-}\bar\phi^{(K)}(\w0,-\v{k}),
\cr}
}
which can be verified to satisfy the boundary condition \bcG.

\appendix{B}{The Boundary-to-Bulk Spinor Propagator}

In this appendix, we will review
the boundary-to-bulk propagator of spinor fields which has been derived in
\refs{\MucVisS,\HenSfe}. 
As in the scalar case \Witten, one can obtain the boundary-to-bulk 
spinor propagator by performing the inversion to 
the solution of the Dirac equation, 
which turns out to depend only on $z_0$.
As pointed out in \HenSfe, upon the inversion of the solution, 
we need to perform the local Lorentz transformation to preserve the 
gauge-fixing condition of the vielbein.
Instead of this procedure,
we will give an alternative derivation of 
the boundary-to-bulk propagator.
The inversion of the Dirac operator is given by
\eqn\conjUD{
\slash{D}(\hat{z})=-U(z)^{-1}\slash{D}(z)U(z)+\hf\Gamma_0,
}
where $U(z)$ is defined in \Killing. This relation follows from
\eqn\U{\eqalign{
&D_{\mu}U(z)=\hf \Gamma_{\mu}U(z)\Gamma_0,
\cr
&U(z)^{-1}\Gamma_{\mu}U(z)=-J_{\mu\nu}(z)\Gamma_{\nu},
\cr}
}
where $J_{\mu\nu}(z)=\delta_{\mu\nu}-2{z_{\mu}z_{\nu}\over z^2}$
 is the conformal Jacobian defined by
\eqn\defJ{
{\d \hat{z}_{\mu}\over \d z_{\nu}}={1\over z^2}J_{\mu\nu}(z).
}
From \conjUD, the Dirac operator in the coordinate $z$ 
can be written as
\eqn\DiracU{
\slash{D}(z)-m=-U(z)\Big[\slash{D}(\hat{z})+m-\hf\Gamma_0\Big]U(z)^{-1}.
}
Using this relation, we can easily see that
\eqn\Sim{
\Sigma=U(z)\hat{z}_0^{{d\over 2}-m\Gamma_0+\hf}
}
satisfies the Dirac equation $(\slash{D}(z)-m)\Sigma=0$.
Since we impose on the boundary-to-bulk propagator of spinor field 
$\Sigma$ the boundary condition that $\Sigma=0$ 
on the boundary $z_0=0$, it should be projected on $\Gamma_0=-1$. 

The boundary-to-bulk propagator can also be  obtained as the limit
of the bulk-to-bulk propagator with one point approaching the boundary.
From \spinprop, we can see that
\eqn\StoSigma{
S(z;\epsilon,\v{x})\rightarrow -\epsilon^{\Delta_{+}-\half}
\Sigma_{\Delta_{+}}(z,\v{x})
}
in the limit $\epsilon\rightarrow 0$. Note that a similar property for
the scalar propagator is used in \KleWit\ to relate 
the boundary behavior of the bulk field and  
the expectation value of the corresponding operator in the boundary theory.

\listrefs

\end